# Quantum Enhanced Correlation Matrix Memories via States Orthogonalisation

Mario Mastriani, and Marcelo Naiouf

*Abstract*—This paper introduces a Quantum Correlation Matrix Memory (QCMM) and Enhanced QCMM (EQCMM), which are useful to work with quantum memories. A version of classical Gram-Schmidt orthogonalisation process in Dirac notation (called Quantum Orthogonalisation Process: QOP) is presented to convert a non-orthonormal quantum basis, i.e., a set of non-orthonormal quantum vectors (called qudits) to an orthonormal quantum basis, i.e., a set of orthonormal quantum qudits. This work shows that it is possible to improve the performance of QCMM thanks QOP algorithm. Besides, the EQCMM algorithm has a lot of additional fields of applications, e.g.: Steganography, as a replacement Hopfield Networks, Bi-level image processing, etc. Finally, it is important to mention that the EQCMM is an extremely easy to implement in any firmware.

## I. INTRODUCTION

SINCE it was first proposed by Feynman [1], that quantum mechanics might be more powerful computationally than a classical Turing machine, we have heard a lot of quantum computational networks [3], quantum cellular automata [2], but only a little about quantum neural networks [4]. The possible reason for the omni-penetrating ideas of quantum information processing (QIP) to avoid the field of artificial neural networks (ANN), is the presence of a nonlinear activation function in any ANN. For very similar reason, a need for nonlinear couplings between optical modes was the main obstacle for building a scalable optical QIP system.

It was shown recently [5], that quantum computation on optical modes using only beam splitters, phase shifters, photon sources and photo detectors is possible. Accepting the ideas of [5], we just assume the existence of a qubit

$$|\psi\rangle = \alpha|0\rangle + \beta|1\rangle, \qquad (1)$$

where $|\alpha|^2 + |\beta|^2 = 1$, with the states $|\alpha\rangle$ and $|\beta\rangle$ are understood as different polarization states of light.

Specifically,

$$|0\rangle = \begin{bmatrix} 1 \\ 0 \end{bmatrix}, \qquad |1\rangle = \begin{bmatrix} 0 \\ 1 \end{bmatrix}$$

and

$$\alpha = \cos\left(\frac{\theta}{2}\right), \qquad \beta = \sin\left(\frac{\theta}{2}\right)e^{i\varphi}$$

with

$$0 \leq \theta \leq \pi, \qquad 0 \leq \varphi < 2\pi$$

See the Bloch's Sphere in Fig.1.

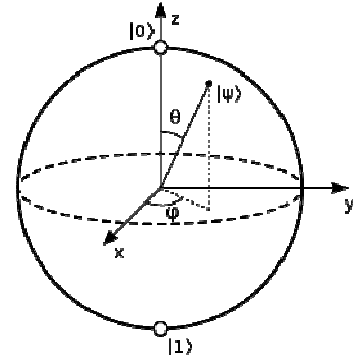

Fig.1 Bloch's Sphere.

On the other hand, a Quantum Associative Memory (QAM) may also be classified as linear or nonlinear, depending on the model adopted for its neurons [6]. In the linear case, the neurons act (to a first approximation) like a linear combiner [7-11]. To be more specific, let the data vectors $|a\rangle$ and $|b\rangle$ denote the stimulus (input) and the response (output) of an associative memory, respectively. In a linear associative memory, the input-out-put relationship is described by

$$|b\rangle = M|a\rangle \qquad (2)$$

where **M** is called the *quantum memory matrix*. The matrix **M** specifies the network connectivity of the QAM. Fig.2 depicts a block-diagram representation of a linear QAM (LQAM). In a nonlinear QAM (NLQAM), on the other hand, we have an input-output relationship of the form

$$|b\rangle = \varphi(M\,;|a\rangle)|a\rangle \qquad (3)$$

where, in general, φ(.;.) is a nonlinear function of the QAM and the input vector.

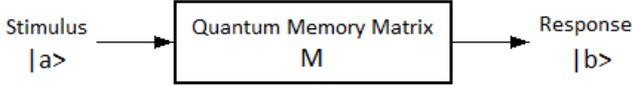

Fig.2 Block diagram of QAM.

Although this form of matrix (in the classical case) was introduced in the 1960's and has been studied intensively since then, see [7-11], and the use of orthogonalisation is not new to improve the storage of these networks, see [9, 10], unfortunately, a CMM version doesn't exist when the *key pattern* $|a_k\rangle$ and the *memorized pattern* $|b_k\rangle$ for all *k* are quantum.

The Quantum Correlation Matrix Memory (QCMM), with the concepts of memory, quantum memory, training and recall (of QCMM) and Enhanced QCMM (EQCMM) is outlined in Section II. In Section III, we discuss briefly the conclusions of the paper.

## II. QUANTUM CORRELATION MATRIX MEMORY

### A. Concept of Memory

Discussion of learning tasks, particularly the task of pattern association, leads us naturally to think about *memory*. In a neurobiological context, memory refers to the relatively enduring neural alterations induced by the interaction of an organism with its environment [6]. Without such a change there can be no memory. Furthermore, for the memory to be useful it must be accessible to the nervous system in order to influence future behavior. However, an activity pattern must initially be stored in memory through a *learning process* [7-14].

Memory, and learning are intricately connected. When a particular activity pattern is learned, it is stored in the brain where it can be recalled later when required. Memory may be divided into "short-term" and "long-term" memory, depending on the retention time [6]. *Short-term memory* refers to a compilation of knowledge representing the "current" state of the environment. Any discrepancies between knowledge stored in short-term memory and a "new" state are used to update the short-term memory. *Long-term memory*, on the other hand, refers to knowledge stored for a long time or permanently.

In this section we study an associative memory that offers the following characteristics:

- The memory is distributed.
- Both the stimulus (key) pattern and the response (stored) pattern of an associative memory consist of data vectors.
- Information is stored in memory by setting up a spatial pattern of neural activities across a large number of neurons.
- Information contained in a stimulus not only determines its storage location in memory but also an address for its retrieval.
- Although neurons do not represent reliable and low-noise computing cells, the memory exhibits a high degree of resistance to noise and damage of a diffusive kind.
- There may be interactions between individual patterns stored in memory. (Otherwise the memory would have to be exceptionally large for it to accommodate the storage of a large number of patterns in perfect isolation from each other). These is therefore the distinct possibility for the memory to make *errors* during the recall process.

In a *distributed memory*, the basic issue of interest is the simultaneous or near-simultaneous activities of many differrent neurons, which are the result of external or internal stimuli. The neural activities form a spatial pattern inside the memory that contains information about the stimuli. The memory is therefor said to perform a distributed mapping that transforms an activity pattern in the input space into another activity pattern in the output space. We may illustrate some important properties of a distributed memory mapping by considering an idealized neural network that consists of two layers of neurons [15-20].

### B. Quantum Memory

In the following mathematical analysis, the quantum neural networks are assumed to be linear. The implication of this assumption is that each neuron acts as a linear combiner. To proceed with the analysis suppose that an activity pattern $|x_k\rangle$ occurs in the input layer of the network and that an activity pattern $|y_k\rangle$ occurs simultaneously in the output layer. The issue we wish to consider here is that of learning form the association between the patterns $|x_k\rangle$ and $|y_k\rangle$. The patterns $|x_k\rangle$ and $|y_k\rangle$ are represented by vectors, written in their expanded forms as:

$$|x_k\rangle = \big[|x_{k1}\rangle, |x_{k2}\rangle, \ldots, |x_{km}\rangle\big]^T$$

and

$$|y_k\rangle = \big[|y_{k1}\rangle, |y_{k2}\rangle, \ldots, |y_{km}\rangle\big]^T$$

From here on we refer to *m* as *network dimensionality* [21]. The elements of both $|x_k\rangle$ and $|y_k\rangle$ can assume positive and negative values.

With this network assumed to be linear, the association of key vector $|x_k\rangle$ with memorized vector $|y_k\rangle$ may be described in matrix form as:

$$|y_k\rangle = W(k)|x_k\rangle, \quad k = 1, 2, \ldots, q \qquad (4)$$

To developed a detailed description of the Eq.(4), we have

$$|y_{ki}\rangle = \sum_{j=1}^{m} w_{ij}(k)|x_{kj}\rangle, \quad i = 1, 2, \ldots, m \qquad (5)$$

where the $w_{ij}(k)$, $j = 1, 2, \ldots, m$ are the synaptic weights of neuron $i$ corresponding to the $k$th pair of associated patterns. Using matrix notation, we may express $y_{ki}$ in the equivalent form

$$|y_{ki}\rangle = [w_{i1}(k), w_{i2}(k), \ldots, w_{im}(k)] \begin{bmatrix} |x_{k1}\rangle \\ |x_{k2}\rangle \\ \vdots \\ |x_{km}\rangle \end{bmatrix}, i = 1, 2, \ldots, m \quad (6)$$

We may define an *m*-by-*m memory matrix* that describes the summation of the weight matrices for the entire set of pattern associations as follows:

$$M = \sum_{k=1}^{q} w(k) \quad (7)$$

The definition of the quantum memory matrix given in Eq.(6) may be restructured in the form of a recursion as shown by

$$M_k = M_{k-1} + W(k), \quad k = 1, 2, \ldots, q \quad (8)$$

where the initial value $M_0$ is zero (i.e., the synaptic weights in the memory are all initially zero), and the final value $M_q$ is identically equal to M as defined in Eq.(7). According to the recursive formula of Eq.(8), the term $M_{k-1}$ is the old value of the memory matrix resulting from ($k$-1) pattern associations, and $M_k$ is the updated value in light of the increment $W(k)$ produced by the $k$th association. Note, however, that when $W(k)$ is added to $M_{k-1}$, the increment $W(k)$ loses its distinct identity among the mixture of contributions that form $M_k$. In spite of the synaptic mixing of different associations, information about the stimuli may not have been lost, as demonstrated in the sequel. Notice also that as the number $q$ of stored patterns increases, the influence of a new pattern on the memory as a whole is progressively reduced.

C. *Quantum Correlation Matrix Memory*

Suppose that the associative memory has learned the memory matrix *M* through the associations of key and memorized patterns described by $|x_k\rangle \to |y_k\rangle$, where $k = 1, 2, \ldots, q$. We may postulate $\hat{M}$, denoting an estimate of the memory matrix *M* in terms of these patterns [6]:

$$\hat{M} = \sum_{k=1}^{q} |y_k\rangle\langle x_k| \quad (9)$$

The term $|y_k\rangle\langle x_k|$ represents the *outer product* of the key pattern $|x_k\rangle$ and the memorized pattern $|y_k\rangle$. This outer product is an *estimate* of the weight matrix W(k) that maps the output pattern $|y_k\rangle$ onto the input pattern $|x_k\rangle$. Since the pattern $|x_k\rangle$ and $|y_k\rangle$ are both *m*-by-1 vectors by assumption, it follows that their output product $|y_k\rangle\langle x_k|$, and there-fore the estimate $\hat{M}$, is an *m*-by-*m* matrix.

On the other hand, Eq.(9) may be reformulated in the equivalent form

$$\hat{M} = \begin{bmatrix} \langle y_1| \\ \langle y_2| \\ \vdots \\ \langle y_q| \end{bmatrix} [|x_1\rangle, |x_2\rangle, \ldots, |x_q\rangle] \quad (10)$$

$$= |Y\rangle\langle X|$$

where

$$\langle X| = [|x_1\rangle, |x_2\rangle, \ldots, |x_q\rangle] \quad (11)$$

and

$$\langle Y| = [\langle y_1|, \langle y_2|, \ldots, \langle y_q|] \quad (12)$$

The matrix X is an *m*-by-*q* matrix composed of the entire set of key patterns used in the learning process; it is called the *key matrix*. The matrix Y is an *m*-by-*q* matrix composed of the corresponding set of memorized patterns; it is called the *memorized matrix*.

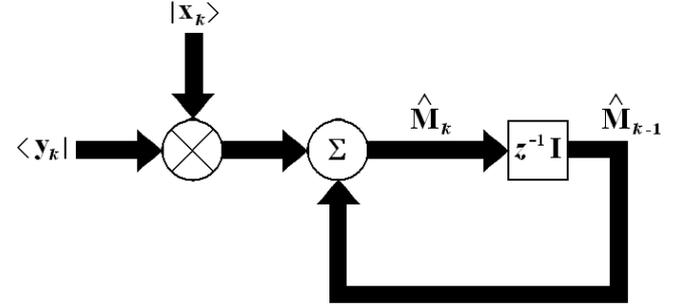

Fig.3 Signal-flow graph representation of Eq.(13).

Equation (10) may also be restructured in the form of a recursion as follows:

$$\hat{M}_k = \hat{M}_{k-1} + |y_k\rangle\langle x_k|, \quad k = 1, 2, \ldots, q \quad (13)$$

A signal-flow graph representation of this recursion is depicted in Fig.(3). According to this signal-flow graph and the recursive formula of Eq.(13), the matrix $\hat{M}_{k-1}$ represents an old estimate of the memory matrix; and $\hat{M}_k$ represents its updated value in the light of a new association performed by the memory on the patterns $|x_k\rangle$ and $|y_k\rangle$. Comparing the recursion of Eq.(13) with that of Eq.(8), we see that the outer product $|y_k\rangle\langle x_k|$ represents an estimate of the weight matrix W(k) corresponding to the $k$th association of key and memorized patterns, $|x_k\rangle$ and $|y_k\rangle$.

## D. Recall

The fundamental problem posed by the use of an associative memory is the address and recall of patterns stored in memory. To explain one aspect of this problem, let $\hat{M}$ denote the memory matrix of an associative memory, which has been completely learned through its exposure to $q$ pattern associations in accordance with Eq.(9). Let a key pattern $|x_j\rangle$ be picked at random and reapplied as *stimulus* to the memory, yielding the *response*

$$|y\rangle = \hat{M}\, x_j \qquad (14)$$

Substituting Eq.(9) in (14), we get

$$|y\rangle = \sum_{k=1}^{m} |y_k\rangle \langle x_k | x_j \rangle$$
$$= \sum_{k=1}^{m} \langle x_k | x_j \rangle |y_k\rangle \qquad (15)$$

Where, in the second line, it is recognized that $\langle x_k | x_j \rangle$ is a scalar equal to the *inner product* of the key vectors $|x_k\rangle$ and $|x_j\rangle$. We may rewrite Eq.(15) as

$$|y\rangle = (\langle x_j | x_j \rangle) |y_j\rangle + \sum_{\substack{k=1 \\ k \neq j}}^{m} (\langle x_k | x_j \rangle) |y_k\rangle \qquad (16)$$

Let each of the key patterns $|x_1\rangle, |x_2\rangle, \ldots, |x_q\rangle$ be normalized to have unit energy; that is,

$$E_k = \sum_{l=1}^{m} |x_{kl}\rangle|x_{kl}\rangle$$
$$= \langle x_k | x_k \rangle$$
$$= 1, \qquad k = 1, 2, \ldots, q \qquad (17)$$

Accordingly, we say simplify the response of the memory to the stimulus (key pattern) $|x_j\rangle$ as

$$|y\rangle = |y_j\rangle + |v_j\rangle \qquad (18)$$

where

$$|v_j\rangle = \sum_{\substack{k=1 \\ k \neq j}}^{m} \langle x_k | x_j \rangle |y_k\rangle \qquad (19)$$

The first term on the right-hand side of Eq.(18) represents the "desired" response $|y_j\rangle$; it may therefore be viewed as the "signal" component of the actual response $|y\rangle$. The second term $|v_j\rangle$ is a "noise vector" that arises because of the *crosstalk* between the key vector $|x_j\rangle$ and all the other key vectors stored in memory. The noise vector $|v_j\rangle$ is responsible for making errors on recall.

In the context of a linear signal space, we may define the *cosine of the angle* between a pair of vectors $|x_j\rangle$ and $|x_k\rangle$ as the inner product of $|x_j\rangle$ and $|x_k\rangle$ divided by the product of their individual Euclidean norms or lengths as shown by

$$\cos(|x_k\rangle, |x_j\rangle) = \frac{\langle x_k | x_j \rangle}{\||x_k\rangle\| \||x_j\rangle\|} \qquad (20)$$

The symbol $\||x_k\rangle\|$ signifies the Euclidean norm of vector $|x_k\rangle$, defined as the square root of the energy of $|x_k\rangle$:

$$\||x_k\rangle\| = \sqrt{\langle x_k | x_k \rangle}$$
$$= \sqrt{E_k} \qquad (21)$$

Returning to the situation, note that the key vectors are normalized to have unit energy in accordance with Eq.(17). We may therefore reduce the definition of Eq.(20) to

$$\cos(|x_k\rangle, |x_j\rangle) = \langle x_k | x_j \rangle \qquad (22)$$

We may then redefine the noise vector of Eq.(19) as

$$|v_j\rangle = \sum_{\substack{k=1 \\ k \neq j}}^{m} \cos(|x_k\rangle, |x_j\rangle) |y_k\rangle \qquad (23)$$

We now see that if the key vectors are *orthogonal* (i.e., perpendicular to each other in a Euclidean sense), then

$$\cos(|x_k\rangle, |x_j\rangle) = |0\rangle, \quad k \neq j \qquad (24)$$

and therefore the noise vector $|v_j\rangle$ is identically zero. In such a case, the response $|y\rangle$ equals $|y_j\rangle$. The *memory associates perfectly* if the key vectors from an *orthonormal set*; that is, if they satisfy the following pair of conditions:

$$\langle x_k | x_j \rangle = \begin{cases} |1\rangle, & k = j \\ |0\rangle, & k \neq j \end{cases} \qquad (25)$$

Suppose now that the key vectors do form an orthonormal set, as prescribed in Eq.(25). What is then the limit on the *storage capacity* of the associative memory? Stated in another way, what is the largest number of patterns that can be reliably stored? The answer to this fundamental question lies in the rank of the memory matrix $\hat{M}$. The rank of a matrix is defined as the number of independent columns (rows) of the matrix. That is, if $r$ is the rank of such a rectangular matrix of dimensions $l$-by-$m$, we then have $r \leq \min(l, m)$. In the case of a correlation memory, the memory matrix $\hat{M}$ is an $m$-by-$m$ matrix, where $m$ is the dimensionality of the input space. Hence the rank of the memory matrix *M* is limited by the dimensionality *m*. We may thus formally state that the number of patterns that can be reliably stored in a correlation matrix memory can never exceed the input space dimensionality.

*E. Enhanced Quantum Correlation Matrix Memory*

According to what explained above, the improvement in the memory consists of removing the noise vector $|v_j\rangle$, which arises from the orthogonalisation of the key vector $|x_k\rangle$, which is conducted through the following procedure [22]. Suppose $|x_1\rangle,\ldots,|x_d\rangle$ is a basis set for some vector space *Z* with an inner product. There is a useful method, the Gram-Schmidt procedure, which can be used to produce an orthonormal basis set $|z_1\rangle,\ldots,|z_d\rangle$ for the vector space *Z*. Define $|z_1\rangle \equiv |x_1\rangle / \||x_1\rangle\|$, and for $1 \le k \le d-1$ define $|z_{k+1}\rangle$ inductively by

$$|z_{k+1}\rangle = \frac{|x_{k+1}\rangle - \sum_{i=1}^{k}\langle z_i | x_{k+1}\rangle |z_i\rangle}{\||x_{k+1}\rangle - \sum_{i=1}^{k}\langle z_i | x_{k+1}\rangle |z_i\rangle\|} \qquad (26)$$

This method is called Quantum Orthogonalisation Process (QOP), being that procedure which allows to automatically switch EQCMM QCMM. It is not difficult to verify that the vectors $|z_1\rangle,\ldots,|z_d\rangle$ form an orthonormal set which is also a basis for *Z*. Thus, any finite dimensional vector space of dimension *d* has an orthonormal basis, $|z_1\rangle,\ldots,|z_d\rangle$ [23-30]. Now the input vectors to QCMM are the z instead of the x. Both blocks (QOP and QCMM) constitute Enhanced QCMM (EQCMM), see Fig.4.

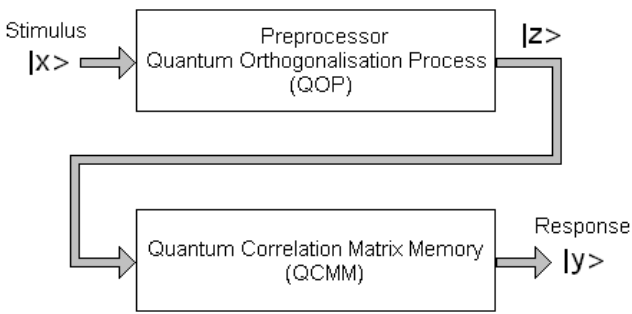

Fig.4 Both blocks (QOP and QCMM) constitute EQCMM.

III. CONCLUSION

We propose a QCMM model and a QOP before that, forming a robust novel QCMM, called EQCMM. QOP removes the key vector noise improvement the recall and the general quantum memory performance.